\documentclass[aps,pre,twocolumn,superscriptaddress]{revtex4}

\usepackage{graphicx}
\usepackage{dcolumn}
\usepackage{bm}
\usepackage{amssymb}
\usepackage{color}

\begin{document}

%\preprint{}

%Title of paper
\title{Two- or three-step assembly of banana-shaped proteins coupled with shape transformation of lipid membranes} 

\author{Hiroshi Noguchi}
\email[]{noguchi@issp.u-tokyo.ac.jp}
\affiliation{
Institute for Solid State Physics, University of Tokyo,
 Kashiwa, Chiba 277-8581, Japan}

%\date{\today}

\begin{abstract}
BAR superfamily proteins have a banana-shaped domain that causes the local bending of lipid membranes.
We study as to how such a local anisotropic curvature induces effective interaction between proteins and
changes the global shape of vesicles and membrane tubes using meshless membrane simulations.
The proteins are modeled as banana-shaped rods strongly adhered to the membrane.
Our study reveals that the rods assemble via two continuous directional phase separations  
unlike a conventional two-dimensional phase separation.
As the rod curvature increases, in the membrane tube
the rods assemble along the azimuthal direction and subsequently along the longitudinal direction
accompanied by shape transformation of the tube.
In the vesicle, in the addition to these two assembly processes,
further increase in the rod curvature induces tubular scaffold formation.
\end{abstract}

%\keywords{}

\maketitle

\section{Introduction}

In living cells, 
membrane shape deformations play a key role in
protein transport, endo/exocytosis, cell motility, and cell division.
In membrane traffic,
proteins are transported by vesicle formation from the donor component and fusion to the target component.
These shape deformations are controlled by various proteins \cite{zimm06,baum10,call13}.
Many of these proteins contain a binding module known as the
 BAR (Bin-Amphiphysin-Rvs) domain, which consists of a banana-shaped dimer \cite{itoh06,masu10,kaba11}.
The BAR domain senses and generates membrane curvature.
The extension of membrane tubes from a giant unilamellar liposome
and absorption of BAR proteins onto tube regions have been experimentally observed \cite{itoh06,sorr12,zhu12,tana13}.
At high density, BAR domains form a cylindrical scaffold \cite{masu10,mim12}.

Since a fluid membrane is isotropic along the membrane surface,
its energy is rotational invariant.
The membrane curvature energy is given by \cite{helf73}
\begin{equation}
F_{\rm {cv}} = \int  \Big\{ \frac{\kappa}{2} (C_1 +
C_2- C_0)^2 + \bar{\kappa} C_1C_2  \Big\} dA,
\label{eq:Helfrich}
\end{equation}
as a second order expansion of the principal curvatures 
 $C_1$ and $C_2$.
The coefficients $\kappa$ and $\bar{\kappa}$
represent the bending rigidity and saddle-splay modulus, respectively,
and $C_0$ denotes the spontaneous curvature.

The tubulation of the membrane can be generated by this isotropic spontaneous curvature $C_0$ \cite{lipo13}. 
However, the BAR domain is banana-shaped, and therefore, it generates an anisotropic curvature.
The anisotropic nature of this curvature has recently attracted considerable attention in terms of
theoretical \cite{kaba11,igli06,wala14} and
 numerical studies \cite{arkh08,simu13,rama12,rama13}.
In this light, phase separation along the longitudinal direction in an axisymmetrical membrane tube \cite{wala14},
 linear aggregation of BAR proteins \cite{simu13},
and discoidal and tubular shapes of vesicles \cite{rama12,rama13} have been reported.
However, our knowledge of the anisotropic effects is still limited to simple geometries and specific situations.
The coupling of phase separation and membrane shapes has not been understood so far.

The aim of this letter is to clarify the membrane-curvature-mediated interactions between BAR-domains
and the difference of these interactions from those generated by the isotropic spontaneous curvature.
We investigate membrane tubes
and vesicles using meshless membrane simulations \cite{nogu09,nogu06,shib11}.
A BAR-domain is modeled as a banana-shaped rod,
and it is assumed to be strongly adsorbed onto the membrane.
In order to focus on the membrane-curvature-mediated interactions,
no direct attractive interaction is considered between the rods.
We  show that the membrane-mediated rod--rod interactions along the parallel and perpendicular directions are quite different
and that they induce directional phase separations.

\section{Simulation Method}

A fluid membrane is represented by a self-assembled one-layer sheet of $N_{\rm {mb}}$  particles.
We employ a spin meshless membrane model \cite{shib11} to account for the membrane spontaneous curvature.
Each particle has an orientational degree of freedom.
Since the details of the meshless membrane model are described in Ref. \cite{shib11},
we only briefly explain the model here.

The particles interact with each other via the potential of their positions ${\bf r}_{i}$ and orientations ${\bf u}_i$ as
\begin{eqnarray}
\frac{U}{k_{\rm B}T} &=\ \ & \hspace{1cm} \sum_{i<j} U_{\rm {rep}}(r_{i,j}) \label{eq:U_all}
               +\varepsilon \sum_{i} U_{\rm {att}}(\rho_i)  \\ \nonumber
&\ \ +& \ \ \frac{k_{\rm{tilt}}}{2} \sum_{i<j} \bigg[ 
( {\bf u}_{i}\cdot \hat{\bf r}_{i,j})^2
 + ({\bf u}_{j}\cdot \hat{\bf r}_{i,j})^2  \bigg] w_{\rm {cv}}(r_{i,j}) \\ \nonumber
&\ \ +&  \frac{k_{\rm {bend}}}{2} \sum_{i<j}  \bigg({\bf u}_{i} - {\bf u}_{j} - C_{\rm {bd}} \hat{\bf r}_{i,j} \bigg)^2 w_{\rm {cv}}(r_{i,j}),
\end{eqnarray} 
where ${\bf r}_{i,j}={\bf r}_{i}-{\bf r}_j$, $r_{i,j}=|{\bf r}_{i,j}|$,
 $\hat{\bf r}_{i,j}={\bf r}_{i,j}/r_{i,j}$, and $k_{\rm B}T$ denotes the thermal energy.
Each particle has an excluded volume with a diameter $\sigma$ via the repulsive potential,
$U_{\rm {rep}}(r)=\exp[-20(r/\sigma-1)]$,
with a cutoff at $r=2.4\sigma$.
Here, we use the potential functions in Ref. \cite{nogu11} instead of Ref. \cite{shib11}
to slightly reduce the numerical costs. 

The second term in Eq. (\ref{eq:U_all}) represents the attractive interaction between the particles.
An attractive multibody potential $U_{\rm {att}}(\rho_i)$ is 
employed to allow the formation of a fluid membrane over wide parameter ranges.
The potential $U_{\rm {att}}(\rho_i)$ is given by
\begin{eqnarray} \label{eq:U_att}
U_{\rm {att}}(\rho_i) = 0.25\ln[1+\exp\{-4(\rho_i-\rho^*)\}]- C,
\end{eqnarray} 
with  $\rho_i= \sum_{j \ne i} f_{\rm {cut}}(r_{i,j})$ and $C= 0.25\ln\{1+\exp(4\rho^*)\}$,
where $f_{\rm {cut}}(r)$ is a $C^{\infty}$ cutoff function.
\begin{equation} \label{eq:cutoff}
f_{\rm {cut}}(r)=\left\{ 
\begin{array}{ll}
\exp\{A(1+\frac{1}{(r/r_{\rm {cut}})^n -1})\}
& (r < r_{\rm {cut}}) \\
0  & (r \ge r_{\rm {cut}}) 
\end{array}
\right.
\end{equation}
with $n=6$, $A=\ln(2) \{(r_{\rm {cut}}/r_{\rm {att}})^n-1\}$,
$r_{\rm {att}}= 1.9\sigma$  $(f_{\rm {cut}}(r_{\rm {att}})=0.5)$, 
and the cutoff radius $r_{\rm {cut}}=2.4\sigma$.
The density $\rho^*=7$ in $U_{\rm {att}}(\rho_i)$ is the characteristic density.
For $\rho_i < \rho^*-1$,
$U_{\rm {att}}(\rho_i)$ acts as a pairwise attractive potential 
while
it approaches a constant value for $\rho_i > \rho^*+1$.
The third and fourth terms in Eq.~(\ref{eq:U_all}) are
discretized versions of the
tilt and bending potentials of the tilt model \cite{hamm98}, respectively.
A smoothly truncated Gaussian function~\cite{nogu06} 
is employed as the weight function 
\begin{equation} \label{eq:wcv}
w_{\rm {cv}}(r)=\left\{ 
\begin{array}{ll}
\exp (\frac{(r/r_{\rm {ga}})^2}{(r/r_{\rm {cc}})^n -1})
& (r < r_{\rm {cc}}) \\
0  & (r \ge r_{\rm {cc}}) 
\end{array}
\right.
\end{equation}
with  $n=4$, $r_{\rm {ga}}=1.5\sigma$, and $r_{\rm {cc}}=3\sigma$.

In this study,
we use   $\varepsilon=5$, and $k_{\rm{tilt}}=k_{\rm {bend}}=10$.
The membrane is in a fluid phase and has
 typical values of the mechanical properties for lipid membranes:
the bending rigidity $\kappa/k_{\rm B}T=15 \pm 1$,
the area of the tensionless membrane per particle $a_0/\sigma^2=1.2778\pm 0.0002$,
the area compression modulus $K_A\sigma^2/k_{\rm B}T=83.1 \pm 0.4$,
and
the line tension of the membrane edge $\Gamma\sigma/k_{\rm B}T= 5.73 \pm 0.04$.
The spontaneous curvature $C_0$ of the membrane is 
given by $C_0\sigma= C_{\rm {bd}}/2$ \cite{shib11}.

The protein rod is modeled as a linear chain of $N_{\rm {sg}}$ membrane particles.
The spontaneous curvature along the rod is denoted as $C_{\rm {rod}}$.
The membrane particles are connected by a bond potential $U_{\rm {rbond}}/k_{\rm B}T = (k_{\rm {rbond}}/2\sigma^2)(r_{i+1,i}-l_{\rm rod})^2$.
The bending potential is given by $U_{\rm {rbend}}/k_{\rm B}T = (k_{\rm {rbend}}/2)(\hat{\bf r}_{i+1,i}\cdot\hat{\bf r}_{i,i-1}- C_{\rm r})^2$,
where $C_{\rm r}=1- (C_{\rm {rod}}l_{\rm rod})^2/2$.
We use $k_{\rm {rbond}}=40$, $k_{\rm {rbend}}=4000$, and $l_{\rm rod}=1.15\sigma$.
The membrane potential parameters between neighboring particles in the rods are modified as
$k_{\rm{tilt}}=k_{\rm {bend}}=40$ and $C_{\rm {bd}}=2C_{\rm {rod}}\sigma$
in order to ensure bending of the rod along the normal to the membrane surface.

We use $N_{\rm {sg}}=10$, which corresponds to the typical aspect ratio of BAR domains.
The BAR domain width is around $2$ nm and the length ranges from $13$ to $27$ nm \cite{masu10}.
Two membrane geometries are investigated:
 a membrane tube of length $L_z=48\sigma$ with a periodic boundary in the $z$ direction
and a spherical vesicle.
For both conditions, the total number of particles is fixed as $N_{\rm {mb}}=2400$.
The volume fraction $\phi_{\rm {rod}}=N_{\rm {rod}}N_{\rm {sg}}/N_{\rm {mb}}$ of rods is varied,
where $N_{\rm {rod}}$ is the number of rods.
Replica exchange molecular dynamics \cite{huku96,okam04} with $128$ replicas is used to obtain the thermal equilibrium states.
The error bars are estimated from four independent runs.
A Langevin thermostat is used to maintain the temperature \cite{shib11,nogu11}.
The results are displayed with lengths normalized by the tube radius $R_{\rm cyl}= 9.89\sigma$  and  vesicle radius $R_{\rm ves}= 15.4\sigma$ 
in the absence of protein rods ($\phi_{\rm {rod}}=0$).
To display the membrane conformation,
its center of mass is set to the origin of the coordinates and
the conformation is rotated to make the eigenvectors of the gyration tensor orient along
 the $x$ and $y$ directions with $\langle x^2\rangle \ge \langle y^2\rangle$ for the membrane tube
and along the $x$, $y$, and $z$ directions with $\langle x^2\rangle \ge \langle y^2\rangle \ge \langle z^2\rangle$ for the vesicle.
For the vesicle, the direction of the $x$ axis is chosen
as the furthest particle from the origin along the $x$ axis has a positive value of the $x$ coordinate.

\begin{figure}
\includegraphics{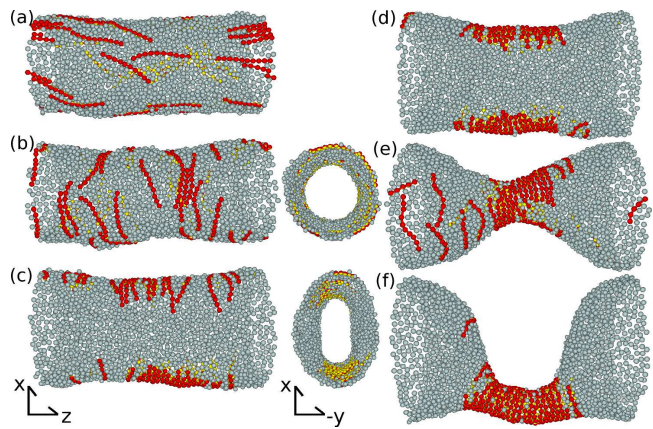}
\caption{
Snapshots of a membrane tube with protein rods at $\phi_{\rm {rod}}=0.167$.
(a) $C_0=0$ and $C_{\rm {rod}}=0$.
(b)  $C_0=0$ and $C_{\rm {rod}}R_{\rm {cyl}}=2$.
(c)  $C_0=0$ and $C_{\rm {rod}}R_{\rm {cyl}}=3$.
(d)  $C_0=0$ and $C_{\rm {rod}}R_{\rm {cyl}}=3.7$.
(e)  $C_0R_{\rm {cyl}}=0.455$ and $C_{\rm {rod}}R_{\rm {cyl}}=3$.
(f)  $C_0R_{\rm {cyl}}=0.455$ and $C_{\rm {rod}}R_{\rm {cyl}}=3.7$.
A protein rod is displayed as
a chain of spheres whose halves are colored
in dark gray (red) and in light gray (yellow).
The orientation vector lies along the direction from the 
light (yellow) to dark gray (red) hemispheres.
Light gray (light blue) particles represent
membrane particles.
The right-side views from the $z$ direction are also shown in (b) and (c).
}
\label{fig:snap_cyl}
\end{figure}

\begin{figure}
\includegraphics{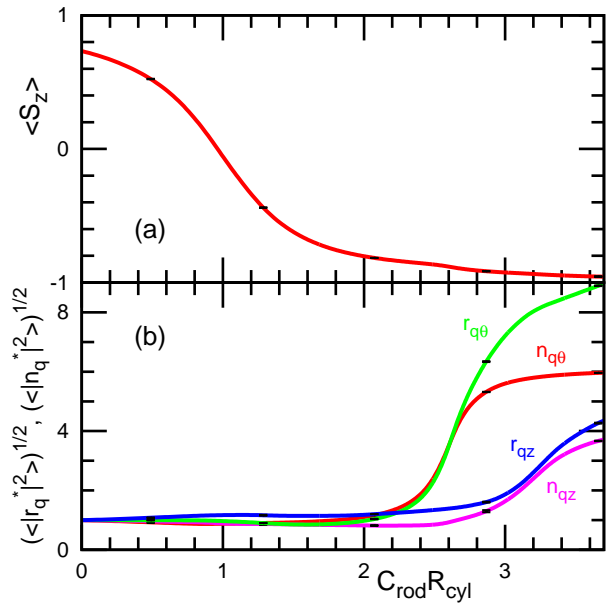}
\caption{
\label{fig:l48c0}
Rod curvature $C_{\rm {rod}}$ dependence of  
(a) the orientation degree $S_z$ and (b) amplitudes of shape deformation and rod densities at  $\phi_{\rm {rod}}=0.167$ and  $C_0=0$.
The amplitudes of the lowest Fourier mode along azimuthal ($\theta$) and longitudinal ($z$) directions
are calculated for the membrane shape ($r_{q\theta}$ and $r_{qz}$) and densities ($n_{q\theta}$ and $n_{qz}$) 
of the center of mass of the protein rods.
The Fourier amplitudes are normalized by those at $C_{\rm {rod}}=0$ (denoted by $*$).
Error bars are displayed at several data points.
}
\end{figure}

\begin{figure}
\includegraphics{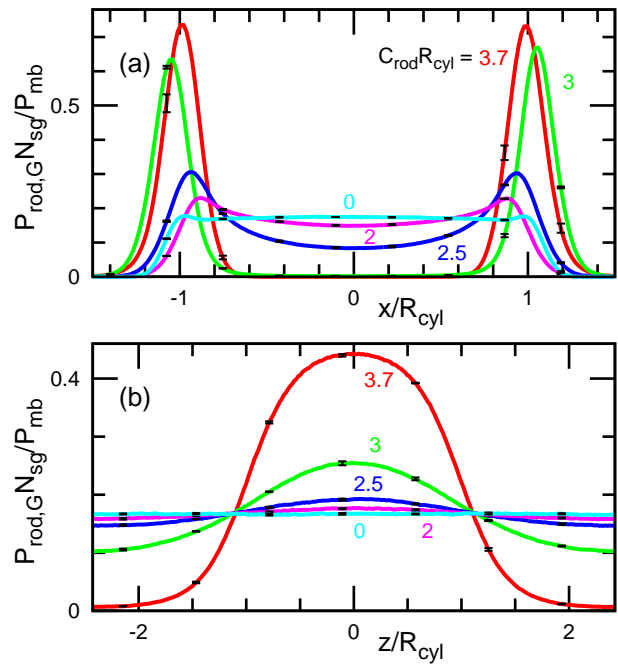}
\caption{
\label{fig:hiscyl}
Probability distribution of the center of mass of the protein rods in the membrane tube along (a) the $x$ and (b) $z$ directions
at $\phi_{\rm {rod}}=0.167$ and  $C_0=0$ for $C_{\rm {rod}}R_{\rm {cyl}}=0$, $2$, $2.5$, $3$, and $3.7$.
The origin of the $z$ coordinate is set at the narrowest position of the membrane determined by the lowest Fourier mode.
Error bars are displayed at several data points.
}
\end{figure}

\section{Membrane Tube}

Figure \ref{fig:snap_cyl} shows snapshots of the membrane tube at $\phi_{\rm {rod}}=0.167$.
Straight rods with $C_{\rm {rod}}=0$ are randomly distributed on the membrane with orientation along the axial ($z$) direction.
As $C_{\rm {rod}}$ increases, the rods rotate into the azimuthal direction
and the orientational order parameter $S_z = (1/N_{\rm rod})\sum_i ( 2{s_{i,z}}^2-1 )$
decreases [see Fig. \ref{fig:l48c0}(a)],
where $s_{i,z}$ is the $z$ component of the orientation vector of the $i$-th rod. 
With further increase in $C_{\rm {rod}}$, the tube transforms from a circular to elliptic cylinder,
and the rods accumulate at the edges of the ellipse [see Fig. \ref{fig:snap_cyl}(c)].
With even further increase, the rods also assemble along the $z$ direction  [see Fig. \ref{fig:snap_cyl}(d)].
As the spontaneous curvature $C_0$ of the membrane increases,
the rods form a narrow cylindrical scaffold [see Figs. \ref{fig:snap_cyl}(e) and (f)].
We note that this scaffold formation is also obtained at $C_0=0$ for longer tubes 
with $L_z=96\sigma$ and $N_{\rm mb}=4800$ and for narrower tubes with $L_z=64\sigma$ and $N_{\rm mb}=2400$.
Since the two-step phase separation is observed more clearly with short tubes,
the present condition ( $L_z=48\sigma$ and $N_{\rm mb}=2400$) is used in this study.

\begin{figure}
\includegraphics{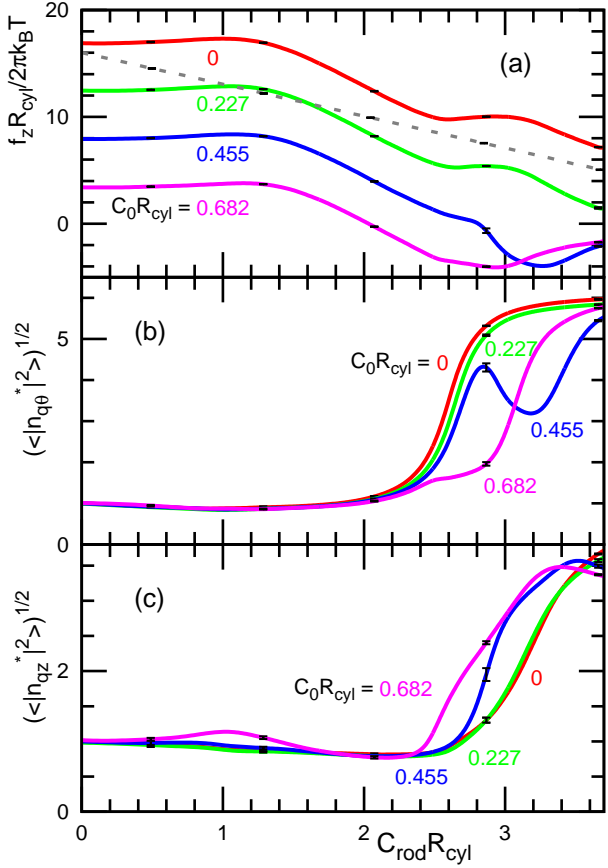}
\caption{
\label{fig:l48cs}
Rod curvature $C_{\rm {rod}}$ dependence of  
(a) the axial force $f_z$ and Fourier amplitudes of rod densities along (b) azimuthal ($n_{q\theta}$) 
and (c) longitudinal ($n_{qz}$) directions  at $\phi_{\rm {rod}}=0.167$ for  $C_0R_{\rm {cyl}}=0$, $0.227$, $0.455$, and $0.682$.
The gray dashed line in (a) represents data for the membrane mixed with particles of isotropic spontaneous curvature.
Error bars are displayed at several data points.
}
\end{figure}
The phase separation between high and low concentrations of the rods is split into two steps along the azimuthal and longitudinal directions.
This phase separation is completely different from conventional phase separations in the two-dimensional space.
We further investigate this two-step phase separation quantitatively.
The amplitudes of the lowest Fourier modes and distribution of the rod positions
are shown in Figs. \ref{fig:l48c0}(b) and \ref{fig:hiscyl}, respectively.
The Fourier modes of the membrane shape are given by  $r_{qz}= (1/N_{\rm mb})\sum_i r_i \exp(-2\pi z_i {\rm i}/L_z)$
and $r_{q\theta}= (1/N_{\rm mb})\sum_i r_i \exp(-2 \theta_i {\rm i})$
where $r_i^2 = x_i^2+ y_i^2$ and $\theta_i=\tan^{-1}(x_i/y_i)$.
The amplitudes of the membrane shape $r_{q\theta}$ and rod density $n_{q\theta}$ 
along the  azimuthal ($\theta$) direction increase together,  
and subsequently, the amplitudes of $r_{qz}$ and rod density $n_{qz}$ along the longitudinal ($z$) direction increase. 
Thus, membrane deformation and rod assembly simultaneously occur along each direction.
The rods are concentrated at the ends of the membrane along the $x$ direction and the narrowest position of the membrane tube
(see Fig. \ref{fig:hiscyl}).
Each phase separation is a continuous transition because each separation is one-dimensional.
Thus, the anisotropy of the spontaneous curvature changes the characteristics of the phase transition.

\begin{figure}
\includegraphics{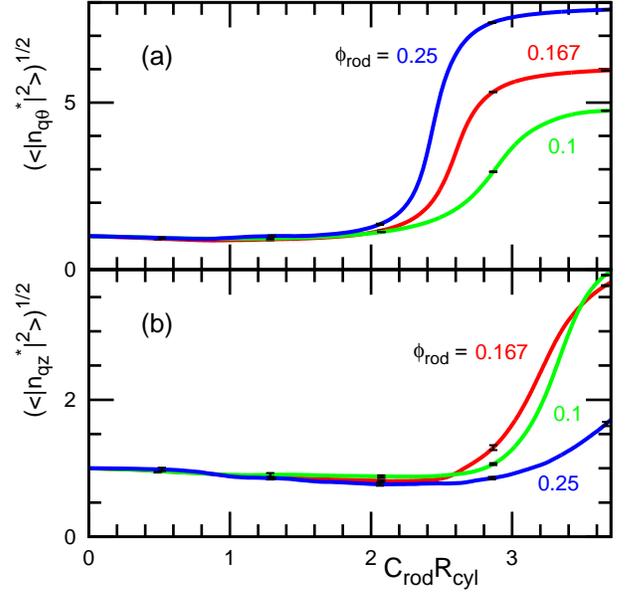}
\caption{
\label{fig:l48n}
Rod curvature $C_{\rm {rod}}$ dependence of  
 Fourier amplitudes of rod densities along (a) azimuthal ($n_{q\theta}$) 
and (b) longitudinal ($n_{qz}$) directions at $C_0=0$ for $\phi_{\rm {rod}}=0.1$, $0.167$, and $0.25$.
Error bars are displayed at several data points.
}
\end{figure}

The anisotropic spontaneous curvature is essential for this two-step phase separation.
When membrane particles with isotropic spontaneous curvatures $C_{\rm rod}$ are mixed in the membrane of $C_0=0$ at the same volume fraction $\phi_{\rm {rod}}=0.167$, the particles remain in a randomly mixed state even at $C_{\rm rod}R_{\rm {cyl}}=3.7$.
A cylindrical tube of homogeneous membranes yields  an axial force
\begin{equation}
f_z = 2\pi\kappa \Big(\frac{1}{R_{\rm cyl}} - C_0\Big), \label{eq:tension}
\end{equation}
since an increase in the axial length leads to a decrease in the cylindrical radius (i.e. a change in the curvature energy) \cite{shib11}.
For the mixture of the membranes with isotropic spontaneous curvatures,
the axial force $f_z$ linearly decreases with $\phi_{\rm {rod}}C_{\rm rod}$ [see the dashed line in Fig. \ref{fig:l48cs}(a)].
Thus, the membrane acts as a homogeneous membrane with the average spontaneous curvature $\phi_{\rm {rod}}C_{\rm rod}$.
In contrast, the membrane with the anisotropic rods exhibits a characteristic change in $f_z$.
For $0\leq C_{\rm rod} \lesssim 1$, the rod orientation changes from the longitudinal to azimuthal direction and
the force $f_z$ is nearly constant [see Fig. \ref{fig:l48cs}(a)].
From $C_{\rm rod} \simeq 1$ until the start of the elliptic deformation,
$f_z$ decreases in a manner similar to the case of a membrane of an isotropic spontaneous curvature.
During the elliptic deformation, $f_z$ is nearly constant.

\begin{figure}
\includegraphics{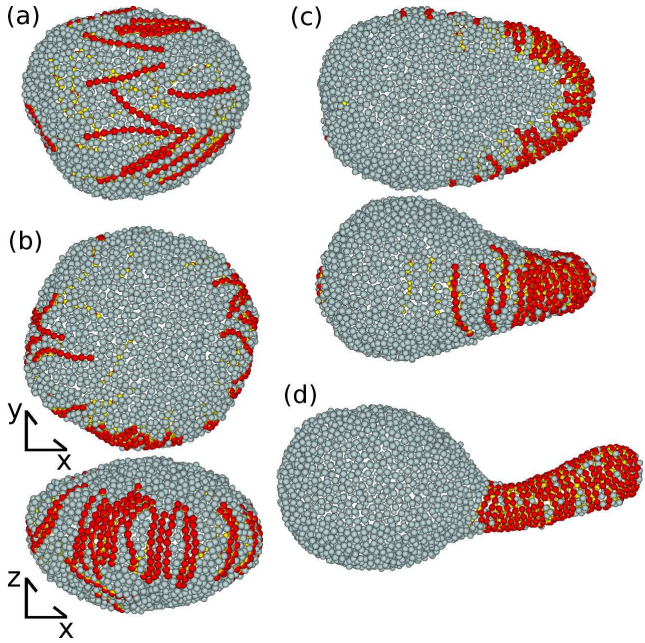}
\caption{
\label{fig:snap_sph}
Snapshots of a vesicle with protein rods at $\phi_{\rm {rod}}=0.167$ and  $C_0R_{\rm {ves}}=0.77$.
(a) $C_{\rm {rod}}=0$.
(b) $C_{\rm {rod}}R_{\rm {ves}}=4.5$.
(c) $C_{\rm {rod}}R_{\rm {ves}}=5.2$.
(d) $C_{\rm {rod}}R_{\rm {ves}}=5.7$.
The top views from the $z$ direction are also shown in (b) and (c).
}
\end{figure}

\begin{figure}
\includegraphics{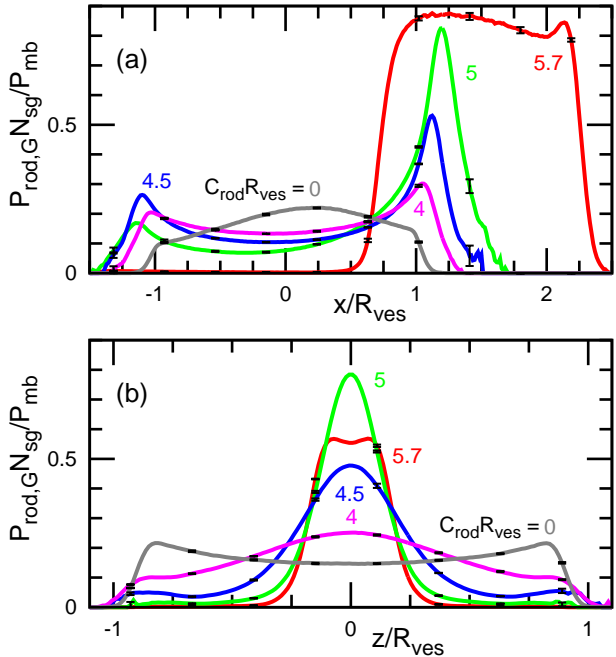}
\caption{
\label{fig:hissph}
Probability distribution of the center of mass of the protein rods along (a) the $x$ and (b) $z$ directions
at $\phi_{\rm {rod}}=0.167$ and  $C_0R_{\rm {ves}}=0.77$ for $C_{\rm {rod}}R_{\rm {ves}}=0$, $4$, $4.5$, $5$, and $5.7$.
Error bars are displayed at several data points.
}
\end{figure}

\begin{figure}
\includegraphics{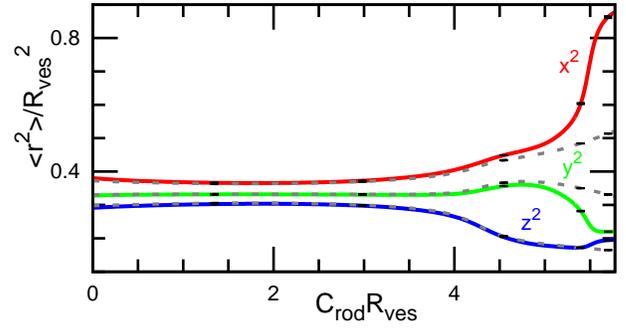}
\caption{
\label{fig:sphxyz}
Three mean eigenvalues of the gyration tensor of the vesicle at $\phi_{\rm {rod}}=0.167$.
The solid and dashed lines represent data for $C_0R_{\rm {ves}}=0.77$ and $0$, respectively.
Error bars are displayed at several data points.
}
\end{figure}

When a finite spontaneous curvature ($C_0\ne 0$) is added to the membrane,
the azimuthal shape deformation and rod assembly are not modified
but the longitudinal deformation is changed [see Figs. \ref{fig:l48cs}(b) and (c)].
This change is induced by the vanishing axial force $f_z$.
With increase in $C_0$,  $f_z$ decreases and the $f_z$ curve in Fig. \ref{fig:l48cs}(a) exhibits a downward shift. 
For $C_0R_{\rm cyl}=0.227$, $f_z$ is always positive and the phase behavior 
is almost identical with that at $C_0=0$ [compare the curves in  Figs. \ref{fig:l48cs}(b) and (c)].
For $C_0R_{\rm cyl}=0.455$, $f_z$ reaches a null value at $C_{\rm rod}R_{\rm cyl}=2.8$, and consequently,
the  longitudinal shape deformation starts before the azimuthal deformation is completed.
For $C_0R_{\rm cyl}=0.682$, $f_z=0$ at $C_{\rm rod}R_{\rm cyl}=2.8$
and the  longitudinal shape deformation starts before the azimuthal deformation.
It is known that negative surface and line tensions induce buckling of flat membranes \cite{nogu11a} and worm-like micelles \cite{otte03}, respectively.
In contrast,  the negative axial force $f_z$ of the membrane tube
induces the longitudinal shape deformation.
For the isotropic membrane, the negative force $f_z$ induces the formation of a small hourglass-shaped neck in the tube 
and subsequent pinch-off into vesicle formation.
In the present anisotropic system, instead, the negative $f_z$ induces longitudinal phase separation
and the resulting narrow cylindrical tube is stabilized by a cylindrical rod scaffold.
The rod scaffold preferentially aligns along the edge of the spherical bud. 
The increase in $n_{q\theta}$ from $C_{\rm rod}R_{\rm cyl}=3.2$ to $3.7$ is caused by
the position change of the narrow tube from the center to the edge [see Figs. \ref{fig:snap_cyl}(e) and (f)].

As the protein volume fraction $\phi_{\rm {rod}}$ increases, 
the azimuthal rod assembly occurs at smaller values of $C_{\rm rod}$,
since the average spontaneous curvature at the elliptic edge increases [see Fig. \ref{fig:l48n}(a)].
In contrast, the longitudinal rod assembly is not sensitive for small values of $\phi_{\rm {rod}}$
but the assembly rate decreases at large values of $\phi_{\rm {rod}}$.
This is because the membrane deformation is suppressed as more than the half region of the edges is occupied by the rods.
Thus,
the protein rods assemble on the membrane tube in the two-step manner
for zero or small values of spontaneous curvature of the membrane and
low volume fraction of the rods.

\section{Vesicle}

The directional phase separations
 also occur in the vesicle (see Figs. \ref{fig:snap_sph} and \ref{fig:hissph}).
With increases in $C_{\rm rod}$,
the rods assemble at the equator of the vesicle 
and the vesicle deforms into an oblate shape.
With further increase, the rods are concentrated at one end 
and vesicle forms a cockscomb-like bump.
These two processes are similar to those in the membrane tube.
With even further increase, 
the rods form a cylindrical scaffold and
the vesicle becomes a tadpole shape.
This additional transformation is not observed in the membrane tube.
The tubule formation is obtained for the vesicle at $C_0R_{\rm ves} \gtrsim 0.9$.

Three principle lengths of the vesicle are shown in Fig. \ref{fig:sphxyz}.
A decrease in $\langle z^2 \rangle$ indicates the oblate formation
with the rod assembly at the equator,
and a sharp increase in $\langle x^2 \rangle$ accompanied by a decrease in $\langle y^2 \rangle$ 
indicates the tubule formation.
The cockscomb formation corresponds to a gradual increase in $\langle x^2 \rangle$ 
(see the dashed line in Fig. \ref{fig:sphxyz})
and a sharp distribution peak of the rod position at $x/R_{\rm ves}\simeq 1$
[see the curve for $C_{\rm rod}R_{\rm ves}=5$ in Fig. \ref{fig:hissph}(a)].

The transition points of the oblate and tubule formations are 
determined by the inflection points of $\langle z^2 \rangle$ and $\langle x^2 \rangle$,
respectively.
The rod curvature at the oblate formation is almost constant; 
$C_{\rm rod}R_{\rm ves}=4.34 \pm 0.06$ and $4.4\pm 0.1$ for $C_0R_{\rm ves}=0$ and $0.85$, respectively.
With increase in $C_0$,
the tubule formation curvature $C_{\rm rod}$ decreases;
$C_{\rm rod}R_{\rm ves}=5.65 \pm 0.02$, $5.48 \pm 0.01$, and $5.33\pm 0.01$ for $C_0R_{\rm ves}=0.69$, $0.77$, and $0.85$, respectively.
At $C_0R_{\rm ves}=1.2$, the oblate phase disappears
and the tubule is formed from a spherical vesicle via a discrete transition.
These $C_0$ dependences are similar to those in the membrane tube.

\section{Conclusions}

We have revealed that the anisotropy of the banana-shaped protein rods induces
splitting of the phase separation into two and three steps
in the membrane tube and vesicle, respectively.
The rods assemble along the orientational (azimuthal) direction first,
and subsequently, they assemble along the perpendicular (longitudinal) direction.
For the vesicle, the tubule is additionally formed at larger rod curvatures.
These directional phase separations are not observed in membranes of isotropic spontaneous curvatures.
Thus, the liquid-crystal nature of the protein rod can significantly change the phase behavior and membrane shapes.

In {\it in vitro} experiments \cite{itoh06,tana13},
the growth of tubules from a vesicle induced by BAR superfamily proteins
has been observed.
In our simulation, before the tubule extension,
the protein rods have already  assembled in the membrane,
and they form a bump of aligned rods.
We speculate that this assembly also occurs in the preceding stage of tubulation in experiments.

It is found that the axial force of the membrane tube is one of the key quantities for 
 the longitudinal membrane deformation and phase separation.
In experiments, the axial force can be measured by optical tweezers \cite{sorr12,zhu12}.
The non-monotonic behavior of the axial force can be used as an indication of phase separation in the tube.

It is considered that the scaffold formation of the BAR proteins is induced by the attractive interaction
between the BAR domains. Our simulations demonstrate that the scaffold can be formed by the membrane-mediated interactions.
The direct attractive interaction can reinforce the protein assembly.
In this study, we assume the protein rods are permanently absorbed on the membrane.
It is known that the BAR domains can be attached and detached from the membrane depending on the membrane curvature.
The effects of this reversible adsorption and rod--rod attractive interactions 
on the phase separation form a topic for future studies.

\begin{acknowledgments}
The replica exchange simulations were
carried out on SGI Altix ICE 8400EX 
at ISSP Supercomputer Center, University of Tokyo. 
This work is supported by KAKENHI (25400425) from
the Ministry of Education, Culture, Sports, Science, and Technology of Japan.
\end{acknowledgments}


\begin{thebibliography}{27}
\expandafter\ifx\csname natexlab\endcsname\relax\def\natexlab#1{#1}\fi
\expandafter\ifx\csname bibnamefont\endcsname\relax
  \def\bibnamefont#1{#1}\fi
\expandafter\ifx\csname bibfnamefont\endcsname\relax
  \def\bibfnamefont#1{#1}\fi
\expandafter\ifx\csname citenamefont\endcsname\relax
  \def\citenamefont#1{#1}\fi
\expandafter\ifx\csname url\endcsname\relax
  \def\url#1{\texttt{#1}}\fi
\expandafter\ifx\csname urlprefix\endcsname\relax\def\urlprefix{URL }\fi
\providecommand{\bibinfo}[2]{#2}
\providecommand{\eprint}[2][]{\url{#2}}

\bibitem[{\citenamefont{Zimmerberg and Kozlov}(2006)}]{zimm06}
\bibinfo{author}{\bibfnamefont{J.}~\bibnamefont{Zimmerberg}} \bibnamefont{and}
  \bibinfo{author}{\bibfnamefont{M.~M.} \bibnamefont{Kozlov}},
  \bibinfo{journal}{Nat.\ Rev.\ Mol.\ Cell\ Biol.}
  \textbf{\bibinfo{volume}{7}}, \bibinfo{pages}{9} (\bibinfo{year}{2006}).

\bibitem[{\citenamefont{Baumgart et~al.}(2010)\citenamefont{Baumgart, Capraro,
  Zhu, and Das}}]{baum10}
\bibinfo{author}{\bibfnamefont{T.}~\bibnamefont{Baumgart}},
  \bibinfo{author}{\bibfnamefont{B.~R.} \bibnamefont{Capraro}},
  \bibinfo{author}{\bibfnamefont{C.}~\bibnamefont{Zhu}}, \bibnamefont{and}
  \bibinfo{author}{\bibfnamefont{S.~L.} \bibnamefont{Das}},
  \bibinfo{journal}{Annu. Rev. Phys. Chem.} \textbf{\bibinfo{volume}{62}},
  \bibinfo{pages}{483} (\bibinfo{year}{2010}).

\bibitem[{\citenamefont{Callan-Jones and Bassereau}(2013)}]{call13}
\bibinfo{author}{\bibfnamefont{A.}~\bibnamefont{Callan-Jones}}
  \bibnamefont{and}
  \bibinfo{author}{\bibfnamefont{P.}~\bibnamefont{Bassereau}},
  \bibinfo{journal}{Curr. Opin. Solid State Mater. Sci.}
  \textbf{\bibinfo{volume}{17}}, \bibinfo{pages}{143} (\bibinfo{year}{2013}).

\bibitem[{\citenamefont{Itoh and Camilli}(2006)}]{itoh06}
\bibinfo{author}{\bibfnamefont{T.}~\bibnamefont{Itoh}} \bibnamefont{and}
  \bibinfo{author}{\bibfnamefont{P.~D.} \bibnamefont{Camilli}},
  \bibinfo{journal}{Biochim.\ Biophys.\ Acta} \textbf{\bibinfo{volume}{1761}},
  \bibinfo{pages}{897} (\bibinfo{year}{2006}).

\bibitem[{\citenamefont{Masuda and Mochizuki}(2010)}]{masu10}
\bibinfo{author}{\bibfnamefont{M.}~\bibnamefont{Masuda}} \bibnamefont{and}
  \bibinfo{author}{\bibfnamefont{N.}~\bibnamefont{Mochizuki}},
  \bibinfo{journal}{Semin. Cell Dev. Biol.} \textbf{\bibinfo{volume}{21}},
  \bibinfo{pages}{391} (\bibinfo{year}{2010}).

\bibitem[{\citenamefont{Kabaso et~al.}(2011)\citenamefont{Kabaso, Gongadze,
  Elter, van Rienen, Gimsa, Kralj-Igli{\v{c}}, and Igli{\v{c}}}}]{kaba11}
\bibinfo{author}{\bibfnamefont{D.}~\bibnamefont{Kabaso}},
  \bibinfo{author}{\bibfnamefont{E.}~\bibnamefont{Gongadze}},
  \bibinfo{author}{\bibfnamefont{P.}~\bibnamefont{Elter}},
  \bibinfo{author}{\bibfnamefont{U.}~\bibnamefont{van Rienen}},
  \bibinfo{author}{\bibfnamefont{J.}~\bibnamefont{Gimsa}},
  \bibinfo{author}{\bibfnamefont{V.}~\bibnamefont{Kralj-Igli{\v{c}}}},
  \bibnamefont{and}
  \bibinfo{author}{\bibfnamefont{A.}~\bibnamefont{Igli{\v{c}}}},
  \bibinfo{journal}{Mini Rev. Med. Chem.} \textbf{\bibinfo{volume}{11}},
  \bibinfo{pages}{272} (\bibinfo{year}{2011}).

\bibitem[{\citenamefont{Sorre et~al.}(2012)\citenamefont{Sorre, Callan-Jones,
  Manzi, Goud, Prost, Bassereau, and Roux}}]{sorr12}
\bibinfo{author}{\bibfnamefont{B.}~\bibnamefont{Sorre}},
  \bibinfo{author}{\bibfnamefont{A.}~\bibnamefont{Callan-Jones}},
  \bibinfo{author}{\bibfnamefont{J.}~\bibnamefont{Manzi}},
  \bibinfo{author}{\bibfnamefont{B.}~\bibnamefont{Goud}},
  \bibinfo{author}{\bibfnamefont{J.}~\bibnamefont{Prost}},
  \bibinfo{author}{\bibfnamefont{P.}~\bibnamefont{Bassereau}},
  \bibnamefont{and} \bibinfo{author}{\bibfnamefont{A.}~\bibnamefont{Roux}},
  \bibinfo{journal}{Proc.\ Natl.\ Acad.\ Sci.\ USA}
  \textbf{\bibinfo{volume}{109}}, \bibinfo{pages}{173} (\bibinfo{year}{2012}).

\bibitem[{\citenamefont{Zhu et~al.}(2012)\citenamefont{Zhu, Das, and
  Baumgart}}]{zhu12}
\bibinfo{author}{\bibfnamefont{C.}~\bibnamefont{Zhu}},
  \bibinfo{author}{\bibfnamefont{S.~L.} \bibnamefont{Das}}, \bibnamefont{and}
  \bibinfo{author}{\bibfnamefont{T.}~\bibnamefont{Baumgart}},
  \bibinfo{journal}{Biophys. J.} \textbf{\bibinfo{volume}{102}},
  \bibinfo{pages}{1837} (\bibinfo{year}{2012}).

\bibitem[{\citenamefont{Tanaka-Takiguchi
  et~al.}(2013)\citenamefont{Tanaka-Takiguchi, Itoh, Tsujita, Yamada,
  Yanagisawa, Fujiwara, Yamamoto, Ichikawa, and Takiguchi}}]{tana13}
\bibinfo{author}{\bibfnamefont{Y.}~\bibnamefont{Tanaka-Takiguchi}},
  \bibinfo{author}{\bibfnamefont{T.}~\bibnamefont{Itoh}},
  \bibinfo{author}{\bibfnamefont{K.}~\bibnamefont{Tsujita}},
  \bibinfo{author}{\bibfnamefont{S.}~\bibnamefont{Yamada}},
  \bibinfo{author}{\bibfnamefont{M.}~\bibnamefont{Yanagisawa}},
  \bibinfo{author}{\bibfnamefont{K.}~\bibnamefont{Fujiwara}},
  \bibinfo{author}{\bibfnamefont{A.}~\bibnamefont{Yamamoto}},
  \bibinfo{author}{\bibfnamefont{M.}~\bibnamefont{Ichikawa}}, \bibnamefont{and}
  \bibinfo{author}{\bibfnamefont{K.}~\bibnamefont{Takiguchi}},
  \bibinfo{journal}{Langmuir} \textbf{\bibinfo{volume}{29}},
  \bibinfo{pages}{328} (\bibinfo{year}{2013}).

\bibitem[{\citenamefont{Mim et~al.}(2012)\citenamefont{Mim, Cui,
  Gawronski-Salerno, Frost, Lyman, Voth, and Unger}}]{mim12}
\bibinfo{author}{\bibfnamefont{C.}~\bibnamefont{Mim}},
  \bibinfo{author}{\bibfnamefont{H.}~\bibnamefont{Cui}},
  \bibinfo{author}{\bibfnamefont{J.~A.} \bibnamefont{Gawronski-Salerno}},
  \bibinfo{author}{\bibfnamefont{A.}~\bibnamefont{Frost}},
  \bibinfo{author}{\bibfnamefont{E.}~\bibnamefont{Lyman}},
  \bibinfo{author}{\bibfnamefont{G.~A.} \bibnamefont{Voth}}, \bibnamefont{and}
  \bibinfo{author}{\bibfnamefont{V.~M.} \bibnamefont{Unger}},
  \bibinfo{journal}{Cell} \textbf{\bibinfo{volume}{149}}, \bibinfo{pages}{137}
  (\bibinfo{year}{2012}).

\bibitem[{\citenamefont{Helfrich}(1973)}]{helf73}
\bibinfo{author}{\bibfnamefont{W.}~\bibnamefont{Helfrich}},
  \bibinfo{journal}{Z.\ Naturforsch} \textbf{\bibinfo{volume}{28c}},
  \bibinfo{pages}{693} (\bibinfo{year}{1973}).

\bibitem[{\citenamefont{Lipowsky}(2013)}]{lipo13}
\bibinfo{author}{\bibfnamefont{R.}~\bibnamefont{Lipowsky}},
  \bibinfo{journal}{Faraday Discuss.} \textbf{\bibinfo{volume}{161}},
  \bibinfo{pages}{305} (\bibinfo{year}{2013}).

\bibitem[{\citenamefont{Igli{\v{c}} et~al.}(2006)\citenamefont{Igli{\v{c}},
  H{\"{a}}gerstrand, Verani{\v{c}}, Plemenita{\v{s}}, and
  Kralj-Igli{\v{c}}}}]{igli06}
\bibinfo{author}{\bibfnamefont{A.}~\bibnamefont{Igli{\v{c}}}},
  \bibinfo{author}{\bibfnamefont{H.}~\bibnamefont{H{\"{a}}gerstrand}},
  \bibinfo{author}{\bibfnamefont{P.}~\bibnamefont{Verani{\v{c}}}},
  \bibinfo{author}{\bibfnamefont{A.}~\bibnamefont{Plemenita{\v{s}}}},
  \bibnamefont{and}
  \bibinfo{author}{\bibfnamefont{V.}~\bibnamefont{Kralj-Igli{\v{c}}}},
  \bibinfo{journal}{J. Theor. Biol.} \textbf{\bibinfo{volume}{240}},
  \bibinfo{pages}{368} (\bibinfo{year}{2006}).

\bibitem[{\citenamefont{Walani et~al.}(2014)\citenamefont{Walani, Torres, and
  Agrawal}}]{wala14}
\bibinfo{author}{\bibfnamefont{N.}~\bibnamefont{Walani}},
  \bibinfo{author}{\bibfnamefont{J.}~\bibnamefont{Torres}}, \bibnamefont{and}
  \bibinfo{author}{\bibfnamefont{A.}~\bibnamefont{Agrawal}},
  \bibinfo{journal}{Phys. Rev. E} \textbf{\bibinfo{volume}{89}},
  \bibinfo{pages}{062715} (\bibinfo{year}{2014}).

\bibitem[{\citenamefont{Arkhipov et~al.}(2008)\citenamefont{Arkhipov, Yin, and
  Schulten}}]{arkh08}
\bibinfo{author}{\bibfnamefont{A.}~\bibnamefont{Arkhipov}},
  \bibinfo{author}{\bibfnamefont{Y.}~\bibnamefont{Yin}}, \bibnamefont{and}
  \bibinfo{author}{\bibfnamefont{K.}~\bibnamefont{Schulten}},
  \bibinfo{journal}{Biophys.\ J.} \textbf{\bibinfo{volume}{95}},
  \bibinfo{pages}{2806} (\bibinfo{year}{2008}).

\bibitem[{\citenamefont{Simunovic et~al.}(2013)\citenamefont{Simunovic,
  Srivastava, and Voth}}]{simu13}
\bibinfo{author}{\bibfnamefont{M.}~\bibnamefont{Simunovic}},
  \bibinfo{author}{\bibfnamefont{A.}~\bibnamefont{Srivastava}},
  \bibnamefont{and} \bibinfo{author}{\bibfnamefont{G.~A.} \bibnamefont{Voth}},
  \bibinfo{journal}{Proc.\ Natl.\ Acad.\ Sci.\ USA}
  \textbf{\bibinfo{volume}{110}}, \bibinfo{pages}{20396}
  (\bibinfo{year}{2013}).

\bibitem[{\citenamefont{Ramakrishnan et~al.}(2012)\citenamefont{Ramakrishnan,
  Ipsen, and {Sunil Kumar}}}]{rama12}
\bibinfo{author}{\bibfnamefont{N.}~\bibnamefont{Ramakrishnan}},
  \bibinfo{author}{\bibfnamefont{J.~H.} \bibnamefont{Ipsen}}, \bibnamefont{and}
  \bibinfo{author}{\bibfnamefont{P.~B.} \bibnamefont{{Sunil Kumar}}},
  \bibinfo{journal}{Soft Matter} \textbf{\bibinfo{volume}{8}},
  \bibinfo{pages}{3058} (\bibinfo{year}{2012}).

\bibitem[{\citenamefont{Ramakrishnan et~al.}(2013)\citenamefont{Ramakrishnan,
  {Sunil Kumar}, and Ipsen}}]{rama13}
\bibinfo{author}{\bibfnamefont{N.}~\bibnamefont{Ramakrishnan}},
  \bibinfo{author}{\bibfnamefont{P.~B.} \bibnamefont{{Sunil Kumar}}},
  \bibnamefont{and} \bibinfo{author}{\bibfnamefont{J.~H.} \bibnamefont{Ipsen}},
  \bibinfo{journal}{Biophys. J.} \textbf{\bibinfo{volume}{104}},
  \bibinfo{pages}{1018} (\bibinfo{year}{2013}).

\bibitem[{\citenamefont{Noguchi}(2009)}]{nogu09}
\bibinfo{author}{\bibfnamefont{H.}~\bibnamefont{Noguchi}},
  \bibinfo{journal}{J.\ Phys.\ Soc.\ Jpn.} \textbf{\bibinfo{volume}{78}},
  \bibinfo{pages}{041007} (\bibinfo{year}{2009}).

\bibitem[{\citenamefont{Noguchi and Gompper}(2006)}]{nogu06}
\bibinfo{author}{\bibfnamefont{H.}~\bibnamefont{Noguchi}} \bibnamefont{and}
  \bibinfo{author}{\bibfnamefont{G.}~\bibnamefont{Gompper}},
  \bibinfo{journal}{Phys.\ Rev.\ E} \textbf{\bibinfo{volume}{73}},
  \bibinfo{pages}{021903} (\bibinfo{year}{2006}).

\bibitem[{\citenamefont{Shiba and Noguchi}(2011)}]{shib11}
\bibinfo{author}{\bibfnamefont{H.}~\bibnamefont{Shiba}} \bibnamefont{and}
  \bibinfo{author}{\bibfnamefont{H.}~\bibnamefont{Noguchi}},
  \bibinfo{journal}{Phys. Rev. E} \textbf{\bibinfo{volume}{84}},
  \bibinfo{pages}{031926} (\bibinfo{year}{2011}).

\bibitem[{\citenamefont{Noguchi}(2011{\natexlab{a}})}]{nogu11}
\bibinfo{author}{\bibfnamefont{H.}~\bibnamefont{Noguchi}},
  \bibinfo{journal}{J.\ Chem.\ Phys.} \textbf{\bibinfo{volume}{134}},
  \bibinfo{pages}{055101} (\bibinfo{year}{2011}{\natexlab{a}}).

\bibitem[{\citenamefont{Hamm and Kozlov}(1998)}]{hamm98}
\bibinfo{author}{\bibfnamefont{M.}~\bibnamefont{Hamm}} \bibnamefont{and}
  \bibinfo{author}{\bibfnamefont{M.~M.} \bibnamefont{Kozlov}},
  \bibinfo{journal}{Eur.\ Phys.\ J.\ B} \textbf{\bibinfo{volume}{6}},
  \bibinfo{pages}{519} (\bibinfo{year}{1998}).

\bibitem[{\citenamefont{Hukushima and Nemoto}(1996)}]{huku96}
\bibinfo{author}{\bibfnamefont{K.}~\bibnamefont{Hukushima}} \bibnamefont{and}
  \bibinfo{author}{\bibfnamefont{K.}~\bibnamefont{Nemoto}},
  \bibinfo{journal}{J.\ Phys.\ Soc.\ Jpn.} \textbf{\bibinfo{volume}{65}},
  \bibinfo{pages}{1604} (\bibinfo{year}{1996}).

\bibitem[{\citenamefont{Okamoto}(2004)}]{okam04}
\bibinfo{author}{\bibfnamefont{Y.}~\bibnamefont{Okamoto}}, \bibinfo{journal}{J.
  Mol. Graph. Model.} \textbf{\bibinfo{volume}{22}}, \bibinfo{pages}{425}
  (\bibinfo{year}{2004}).

\bibitem[{\citenamefont{Noguchi}(2011{\natexlab{b}})}]{nogu11a}
\bibinfo{author}{\bibfnamefont{H.}~\bibnamefont{Noguchi}},
  \bibinfo{journal}{Phys. Rev. E} \textbf{\bibinfo{volume}{83}},
  \bibinfo{pages}{061919} (\bibinfo{year}{2011}{\natexlab{b}}).

\bibitem[{\citenamefont{den Otter et~al.}(2003)\citenamefont{den Otter,
  Shkulipa, and Briels}}]{otte03}
\bibinfo{author}{\bibfnamefont{W.~K.} \bibnamefont{den Otter}},
  \bibinfo{author}{\bibfnamefont{S.~A.} \bibnamefont{Shkulipa}},
  \bibnamefont{and} \bibinfo{author}{\bibfnamefont{W.~J.}
  \bibnamefont{Briels}}, \bibinfo{journal}{J.\ Chem.\ Phys.}
  \textbf{\bibinfo{volume}{119}}, \bibinfo{pages}{2363} (\bibinfo{year}{2003}).

\end{thebibliography}
\end{document}